\documentclass[letter]{aa} 

\usepackage{graphicx}
\usepackage{txfonts}
\usepackage[colorlinks=true,     linkcolor=blue, citecolor=blue, filecolor=blue, urlcolor=blue]{hyperref}
\defcitealias{Quintana2025}{Q25}

\defcitealias{HuntReffert2024}{HR24}

\begin{document}

\title{How many stars form in compact clusters in the local Milky Way?}

\author{Alexis L. Quintana\inst{\ref{inst1}}\thanks{Corresponding author: alexis.quintana@obspm.fr}\and Emily L. Hunt \inst{\ref{inst2}} \and Hanna Parul\inst{\ref{inst1}}} 
\institute{LIRA, Observatoire de Paris, Université PSL, Sorbonne Université, Université Paris Cité, CY Cergy Paris Université, CNRS, 92190 Meudon, France \label{inst1} \and
Max-Planck-Institut f\"ur Astronomie, K\"onigstuhl 17, 69117 Heidelberg, Germany \label{inst2}}

\date{Received 11 July 2025 ; Accepted 14 August 2025}

 
\abstract
   {Two main models coexist for the environment in which stars form. The clustered model stipulates that the bulk of star formation occurs within dense embedded clusters, but only a minority of them survive the residual gas expulsion phase caused by massive stellar feedback unbinding the clusters. On the other hand, the hierarchical model predicts that star formation happens at a range of scales and densities, where open clusters (OCs) only emerge from the densest regions.}   
   {We aim to exploit a recent catalog of compact OCs, corrected for completeness, to obtain an updated estimation of the surface density star formation rate within OCs ($\sum_{\rm SFR,OC}$), which we compare with recent estimates of $\sum_{\rm SFR}$ to determine which model is more likely.}
   {We have applied two methods. The first one consisted of integrating over the power law that was fit for the mass function of the youngest OCs using a MC sampling. The second one consisted of counting the total compact mass within these youngest OCs within 1 kpc, so that the result could be directly compared with local values of $\sum_{\rm SFR}$.}
   {We estimated new $\sum_{\rm SFR,OC}$ values between $736^{+159}_{-176}$ and $875^{+34}_{-35}$ M$_{\odot}$ Myr$^{-1}$ kpc$^{-2}$, depending on the methodology. These results are significantly higher than previous $\sum_{\rm SFR,OC}$ estimates, which we attribute to the incompleteness of past catalogs, and are consistent with the majority ($\geq$ 50 \%) or even the vast majority ($\geq$ 80 \%) of the star formation occurring in initially compact clusters, through comparisons with $\sum_{\rm SFR}$ from the recent literature.}
   {Our new $\sum_{\rm SFR,OC}$ values are consistent with clustered formation being the most dominant mode of star formation.}


 \maketitle
%

\section{Introduction}
\label{intro}

The environment in which stars form remains an open question in astronomy, for which there exist several scenarios. The clustered model from \citet{LadaLada2003} predicts that the majority of stars (70--90\%) emerge from dense, gravitationally bound embedded clusters. Only a fraction ($\sim$5 \%) of these clusters will survive past the dispersion of the molecular gas caused by the feedback from the massive newborn stars, a phenomenon known as residual gas expulsion \citep{Hills,Lada1984,Kroupa2001}. Those that survive will remain as bound open clusters (OCs) for up to 1 Gyr (\citealt{Lada1991}), while those that disperse will be briefly visible as low-density and gravitationally unbound groups named associations, which will expand and dissolve into the Galactic field population of stars after a few tens of millions of years \citep{Brown1997,Wright2020}.

An alternative scenario, known as the hierarchical model, predicts that stars can form over a wide range of densities and scales. In this model, OB associations arise from regions of low-density (and are therefore unbound from birth), whereas bound OCs emerge from the densest regions, implying that only a minority of field stars originate from initially clustered environments \citep{Heyer2001,Elmegreen2008,Kruijssen2012}. It is also possible that the reality lies somewhere between these two models \citep{Wright2020}, partly due to the difference in timescale between the formation of stars (indicated by their isochronal age) and the assembly and/or dissolution of star clusters \citep[e.g.][]{MiretRoig2024}.

The arrival of data from ESA's \textit{Gaia} satellite \citep{Gaia} has enabled the 3D positions and 2D velocities of stars to be mapped with unprecedented accuracy, and can therefore help to constrain one scenario over the other. For instance, \citet{Ward2020} found that the highly substructured velocity field exhibited by their catalog of OB associations is more consistent with the hierarchical scenario. On the other hand, updated memberships of OB associations with more kinematic coherence than their historical counterparts have enabled the detection of more significant expansion signatures compatible with the clustered model \citep[e.g.,][]{Kounkel2018,CantatGaudin2019,QuintanaWright2021}, although certain asymmetrical expansion patterns could also be explained by the hierarchical model \citep{Kruijssen2012}.

Properties of star formation can also be explored with catalogs of OCs. For example, \citet{LamersGieles2006} calculated a star formation surface density rate within compact clusters (hereafter, $\sum_{\rm SFR,OC}$) of 350 M$_{\odot}$ Myr$^{-1}$ kpc$^{-2}$ based on a catalog of OCs within 600 pc \citep{Kharchenko2005} that they believed to be complete at the time. Revisiting this value is particularly worthwhile as many new clusters have been identified thanks to \textit{Gaia} data, including within 600 pc \citep[e.g.,][]{LiuPang2019,CantatGaudin2020,CastroGinard2020}. Notably, \citet{Anders2021} derived a value of  $\sum_{\rm SFR,OC} =250^{+190}_{-130}$ M$_{\odot}$ Myr$^{-1}$ kpc$^{-2}$, based on the catalog of OCs from \citet{CantatGaudin2020} and by assuming a mean initial cluster mass of between $\sim${}300 to 600 M\textsubscript{\sun}. Comparing it with the value of $1600^{+700}_{-400}$ M$_{\odot}$ Myr$^{-1}$ kpc$^{-2}$ from \citet{Mor2019}, derived from a census of stars with $G < 12$ mag in \textit{Gaia} DR2 \citep{GaiaDR2}, they concluded that only $16^{+11}_{-8}$ \% of stars in the solar neighborhood formed in bound OCs. 

Recent measurements warrant a revisit of the fraction of stars that form in compact clusters. Firstly, \citet[][hereafter Q25]{Quintana2025} revised down the local value of $\sum_{\rm SFR}$ to $922^{+133}_{-1}$ M$_{\odot}$ Myr$^{-1}$ kpc$^{-2}$ using their highly complete census of O- and B-type stars within 1 kpc, providing the tightest constraint yet on the local value of $\sum_{\rm SFR}$. Open cluster catalogs have also seen large strides in the past five years. Using \textit{Gaia} DR3 \citep{GaiaDR3}, \citet{HuntReffert2023} have produced the largest ever homogeneous and de-duplicated catalog of star clusters, which includes many new clusters within the solar neighborhood that were not included in the previous bound cluster $\sum_{\rm SFR}$ estimate from \cite{Anders2021}\footnote{\citetalias{HuntReffert2024} includes 744 compact OCs within 1 kpc, hence nearly double the census of 382 OCs in \citet{CantatGaudin2020}.}. Since the method of \cite{HuntReffert2023} also appeared to be sensitive to unbound star clusters (such as OB associations or moving groups), \citet[][hereafter HR24]{HuntReffert2024} created a revised catalog that uses cluster masses and Jacobi radii to divide their clusters into bound clusters (OCs) and unbound clusters. This definition of cluster boundedness does not use stellar velocities, and so many of their `bound' clusters may actually be gravitationally unbound (e.g., as after gas expulsion or other disruptive events); nevertheless, this work comprises a large catalog of self-gravitating clusters with measured ages and masses. We therefore refer to these clusters as “compact”, and further address the issue of boundedness of clusters in Section \ref{caveat}.

In this paper, we exploit this catalog of compact OCs to estimate an updated value of $\sum_{\rm SFR,OC}$ in the local Milky Way, finding contrary to recent studies that a majority of stars originate from initially bound clusters. In Section \ref{SFRcalculation}, we detail the two methods applied to estimate this quantity. In Section \ref{discussion}, we compare our results with literature values and discuss the implications for star formation scenarios.

\section{Surface density star formation rate calculations}
\label{SFRcalculation}

In this section, we describe the two methods we applied to derive $\sum_{\rm SFR,OC}$ based on the catalog of compact OCs from \citetalias{HuntReffert2024}. Since we are looking at the rate at which stars form in compact clusters, we adopt the cluster mass contained within the Jacobi radius (M$_J$) as the cluster mass, as defined in \citetalias{HuntReffert2024}. In both cases, our calculations are based on their completeness correction. This correction depends on cluster mass and distance, with the cluster distribution peaking, respectively, at $\sim$1 and 2.7 kpc for the lowest (50 $\leq$ M$_J$ $<$ 100 M$_{\odot}$) and highest (800 $\leq$ M$_J$ $<$ 1600 M$_{\odot}$ and 1600 $\leq$ M$_J$ $<$ 3200 M$_{\odot}$) mass bins. We restricted our analysis to the OCs with a median age below 10~Myr (i.e., $\log$ t $<$ 7), the youngest age bin defined in Section 5.1 from \citetalias{HuntReffert2024}. We consider the current mass of these OCs (M$_J$) to correspond to their initial mass, and therefore assume mass loss in OCs to be negligible over the first 10~Myrs of their life, which is a reasonable approximation given models of long-term cluster dissolution \citep[e.g.,][]{LamersGieles2006,Almeida2025}. Finally, for both methods, we also calculated  $\sum_{\rm SFR,OC}$ values based on a subset of OCs with an astrometric signal-to-noise ratio greater than 5$\sigma$: this cluster significance test (hereafter CST) is defined in \citet{HuntReffert2021} as the level of separation of cluster members with field stars using a nearest-neighbor method. \citet{HuntReffert2023,HuntReffert2024} defined their high-quality sample of OCs with a second condition (a median CMD class, $Q_{\rm CMD}$, greater than 0.5). Here we do not apply this cut because we focus on the youngest OCs, many of which are still embedded in gas, which therefore broadens their CMD due to differential reddening. The median A$_V$ value of the completeness-corrected sample of compact OCs younger than 10~Myr with a good $Q_{\rm CMD}$ is equal to 1.72 mag, while it is equal to 2.73 mag for those with a bad $Q_{\rm CMD}$.

\begin{table}
	\centering
	\caption{Surface density star formation rates ($\sum_{\rm SFR,OC}$) derived from the completeness-corrected mass function of compact OCs from \citetalias{HuntReffert2024}. \label{SFRvalues}}
	\renewcommand{\arraystretch}{1.5} 
	\begin{tabular}{lcccr} 
		\hline
		Parameter & Value & Units  \\
		\hline
        $\sum_{\rm SFR,OC}$ (power law) & $791^{+162}_{-191}$ & M$_{\odot}$ Myr$^{-1}$ kpc$^{-2}$ \\
        $\sum_{\rm SFR,OC,CST \geq5}$ (power law) &  $736^{+159}_{-176}$ & M$_{\odot}$ Myr$^{-1}$ kpc$^{-2}$ \\
        $\sum_{\rm SFR,OC} /\sum_{\rm SFR}$ & $86^{+14}_{-24}$ \% & \\
        $\sum_{\rm SFR,OC, CST \geq5} /\sum_{\rm SFR}$ & $80^{+17}_{-22}$ \% & \\
        \hline
        $\sum_{\rm SFR,OC}$ (cluster count) &  $875^{+34}_{-35}$ & M$_{\odot}$ Myr$^{-1}$ kpc$^{-2}$ \\
        $\sum_{\rm SFR,OC, CST \geq5}$ (cluster count) & $846^{+34}_{-35}$ & M$_{\odot}$ Myr$^{-1}$ kpc$^{-2}$ \\
        $\sum_{\rm SFR,OC} /\sum_{\rm SFR}$ & $95^{+4}_{-14}$ \% & \\
        $\sum_{\rm SFR,OC, CST \geq5} /\sum_{\rm SFR}$ & $92^{+4}_{-14}$ \% & \\

		\hline
	\end{tabular}
\begin{minipage}{\linewidth} 
\vspace{0.25cm}
Note: The first value corresponds to the $\sum_{\rm SFR,OC}$ derived from the power-law fit of the mass function (Section \ref{powerlaw}), whereas the second value stands for the $\sum_{\rm SFR,OC}$ derived through a cluster count for OCs within 1 kpc (Section \ref{clustercount}). For each value, we have also displayed the ratio compared with the $\sum_{\rm SFR}$ of $922^{+133}_{-1}$ M$_{\odot}$ Myr$^{-1}$ kpc$^{-2}$ from \citetalias{Quintana2025}.
\end{minipage}
\end{table}

\subsection{Derivation from power law}
\label{powerlaw}

To correct the observed numbers of clusters for completeness, \citetalias{HuntReffert2024} used an approach that was based on the expected distribution of OCs within the Milky Way's disk. Because more massive clusters can be detected at higher distances than lower-mass clusters, this correction is volume-dependent. The $\sum_{\rm SFR,OC}$ value we thus derive from the resulting mass function is valid across a broad local region of the Milky Way.

\begin{figure}
    \centering
    \includegraphics[scale=0.35]{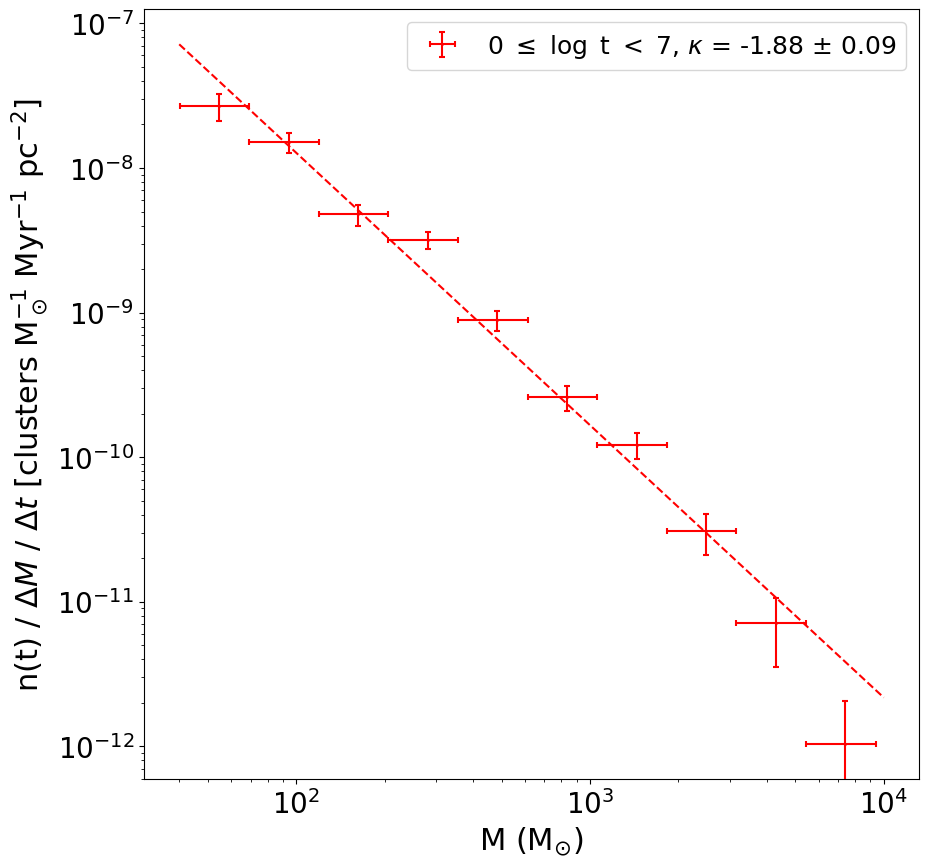}
    \caption{Completeness-corrected mass function for the compact OCs from \citetalias{HuntReffert2024} younger than 10~Myr, including their power-law fit.}
    \label{PowerLawYoungOCs}
\end{figure}

The mass function of OCs younger than 10~Myr was fit with a power law with a slope of $\kappa = -1.88 \, \pm \, 0.09$ in \citetalias{HuntReffert2024}, close to the -2 value from \citet{Krumholz2019}, as is illustrated in Fig. \ref{PowerLawYoungOCs}. We defined a covariance matrix from the fit of the mass function that includes the power-law index alongside the normalization factor. To integrate over the power law, we used a Monte-Carlo (MC) sampling with 1 million random samples drawn from a 2D Gaussian (multivariate normal) distribution. The integration was performed between a mass of 40 M$_{\odot}$ (a fair value for the minimum mass of an OC, \citetalias{HuntReffert2024}) and 10,000 M$_{\odot}$ (the maximum mass of a cluster for which the mass function is well approximated by a power law, \citealt{PortegiesZwart2010}). The result of this process is displayed in Table \ref{SFRvalues}.

\subsection{Derivation from cluster count}
\label{clustercount}

As an additional, simpler method of deriving a $\sum_{\rm SFR,OC}$, we tried simply counting OCs within 1 kpc. We thus further restricted our list of compact OCs to those within $\sqrt{X^2+Y^2} < 1$ kpc, where  $X = d \, \cos(l) \, \cos(b)$ and $Y = d \, \sin(l) \, \cos(b)$ are heliocentric Galactic Cartesian coordinates, in order to enable a more direct comparison with $\sum_{\rm SFR}$ values computed from local catalogs. Since this sample of OCs is complete down to 100 M$_{\odot}$ \citepalias{HuntReffert2024}, this result is thus only valid for clusters above this limit, but this is just slightly above the lower mass limits for 40--60 M$_{\odot}$ for OCs \citep[e.g.,][]{HuntReffert2024,Almeida2025}. We followed the approach from \citetalias{Quintana2025} and derived the $\sum_{\rm SFR,OC}$ within 1 kpc through a “counting” method. To that end, we performed a MC simulation whereby we normally randomized the cluster ages. This allowed us to select the clusters younger than 10~Myr at each step, after which we also randomized the cluster masses within their uncertainties, and summed over the resulting masses of each cluster. We repeated this process over 1 million times, dividing the values by the surface area of our 1~kpc sample and assuming a constant cluster formation rate every million years over the past ten million years. We then adopted the median value as the $\sum_{\rm SFR,OC}$, and the 16\textsuperscript{th} and 84\textsuperscript{th} percentiles as its lower and upper error, respectively. Finally, we found that our value of $\sum_{\rm SFR,OC}$ is relatively insensitive to the choice of cluster age catalog. We indeed derived $\sum_{\rm SFR,OC}$ = 850 $\pm$ 30 M$_{\odot}$ Myr$^{-1}$ kpc$^{-2}$ and $\sum_{\rm SFR,OC, CST \geq5}$ = $835^{+30}_{-31}$ M$_{\odot}$ Myr$^{-1}$ kpc$^{-2}$ using ages from \citet{Cavallo2024}.

\section{Discussion}
\label{discussion}

\subsection{Comparison with literature values}
All of our $\sum_{\rm SFR,OC}$ values in Table \ref{SFRvalues} are noticeably larger than the value of 350 M$_{\odot}$ Myr$^{-1}$ kpc$^{-2}$ obtained in \citet{LamersGieles2006} from a catalog of OCs \citep{Kharchenko2005} within 600 pc with an initial mass between 100 and 30,000 M$_{\odot}$. This result is unsurprising since pre-\emph{Gaia} catalogs of star clusters have been shown to be incomplete \citep[e.g.,][]{CantatGaudin2022,CastroGinard2022,HuntReffert2023}, so $\sum_{\rm SFR,OC}$ values derived from such catalogs are likely to be underestimated. However, our results are also significantly higher than the value of $250^{+190}_{-130}$ M$_{\odot}$ Myr$^{-1}$ kpc$^{-2}$ from \citet{Anders2021}, which exploited the catalog of OCs from \citet{CantatGaudin2020} compiled with \textit{Gaia} DR2. This paper, however, assumed a mean cluster initial mass of $\sim$300-600 M$_{\odot}$, while \citetalias{HuntReffert2024} fit an individual mass for each cluster. Given that OCs have a power-law distribution of initial masses \citep[][see also Fig.~\ref{PowerLawYoungOCs}]{PortegiesZwart2010,Krumholz2019}, which is also reflected in \citetalias{HuntReffert2024}'s catalog, this is likely to be more accurate than assuming a typical mean cluster mass. The median mass of the subsample of compact OCs younger than 10 Myr from \citetalias{HuntReffert2024} is $\sim$130 M$_{\odot}$, thanks to recent discoveries in \emph{Gaia} data of clusters previously too small or low-mass to see. These new low-mass clusters are still consistent with the expected slope of the cluster mass distribution, and many have a high statistical significance that strongly suggests that they are real objects \citepalias{HuntReffert2024}. By contrast, our $\sum_{\rm SFR}$ values are slightly smaller than the lower end of the interval of 1000--3000 M$_{\odot}$ Myr$^{-1}$ kpc$^{-2}$ from \citet{LadaLada2003}, derived from a sample of embedded clusters within 500 pc.

Our $\sum_{\rm SFR,OC}$ values are consistent within the error bars or slightly lower than the value of $922^{+133}_{-1}$ M$_{\odot}$ Myr$^{-1}$ kpc$^{-2}$ from \citetalias{Quintana2025}, based on a \textit{Gaia} DR3 census of O- and B-type stars within 1 kpc, for which they assumed a constant star formation rate over the lifetime of a late B-type star (i.e., $\sim$400 Myr according to the stellar evolutionary models from \citealt{Ekstrom}). They are nonetheless smaller than the value of $1600^{+700}_{-400}$ M$_{\odot}$ Myr$^{-1}$ kpc$^{-2}$ from \citet{Mor2019} based on a \textit{Gaia} DR2 census of stars with $G <$ 12 mag within the Galactic thin disk, with the knowledge that $\sim$94~\% of stars in \textit{Gaia} DR3 with $G <$ 12 mag are contained within $\sqrt{X^2+Y^2} < 1$ kpc using the distances from \citet{BailerEDR3}.

Nevertheless, comparisons of $\sum_{\rm SFR,OC}$ with  \citetalias{Quintana2025} are compatible with the majority of stars forming in initially compact clusters in the solar neighborhood (typically greater than 80 \%, as is displayed in Table \ref{SFRvalues}). While a comparison with \citet{Mor2019} provides a more conservative estimate, the fraction is still above 50 \% within the error bars, a significant increase over the fraction of $16^{+11}_{-8}$ \% from \citet{Anders2021}.

\subsection{Implications for star formation}
Our results suggest that a higher fraction of stars form in compact clusters than previously thought, which is compatible with the clustered formation model from \citet{LadaLada2003}. Disagreement with works that have favored the hierarchical scenario could be attributed to the rapid destruction of OCs\footnote{Here, however, we solely focus on long-term dissolution, as residual gas expulsion can act on a shorter timescale of a few million years to unbind rapidly expanding OCs that still appear clustered according to the Jacobi criterion \citep{Wright2024}; hence, the compact denomination.}. For example, the analysis in \citet{Ward2020} relies upon a list of OB association members dominated by B-type stars \citep{MelnikDambis2020}, whose lifetime can exceed several hundred million years. However, the dissolution time of an OC with an initial mass of 130 M$_{\odot}$ is estimated to be around 93~Myr for clusters surviving residual gas expulsion \citep{Lamers2005,LamersGieles2006,Anders2021}\footnote{\citet{Just2023} and \citet{Almeida2025} estimated a greater value for the long-term dissolution timescale, but their analysis relied upon incomplete catalogs of OCs, as the former exploited data from MWSC \citep{Kharchenko2012,Kharchenko2013} with $\sim$2800 OCs and the latter from \citet{Dias2021} with $\sim$1800 OCs, compared with $\sim$5600 OCs in \citetalias{HuntReffert2024}. Furthermore, neither \citet{Just2023} nor \citet{Almeida2025} have corrected their sample of OCs for selection effects, contrary to \citet{Anders2021} and \citetalias{HuntReffert2024}.}. Likewise, \citet{Quintana2023} identified $\sim$5600 O- and B-type stars in the Auriga region (toward the Galactic anticenter), but only $\sim$10\% ended up grouped into OB associations. They notably attributed this low fraction to their sample being dominated by late B-type stars. In \citet{Quintana2025b}, the fraction of early-type stars centered around Cas OB5 found in star clusters is even lower ($\sim$2\%), since it extends down to the A5 type whose lifetime is in the range of 1.5--2 Gyr \citep[e.g.,][]{Ekstrom}. In those cases, a likely scenario is that these early-type stars have survived past the dissolution of their natal cluster or association to become field stars.

With modern data, it should now be possible to trace the post-disruption stage of clustered star formation in more detail than ever before, which could also help to provide an intermediate constraint on the fraction of stars that form in compact clusters. Within a 250$^3$ pc$^3$ volume near to Sco-Cen, \citet{Ratzenbock2025} found that so-called “disk streams” of stars from disrupted clusters exist at a surface density of 160 kpc\textsuperscript{-2}, surpassing previous estimates of the number of disk streams by one to two orders of magnitude. These streams should be more widely detected in the future and would then offer another picture of the timescale of cluster disruption, helping to explain whether older OB associations (on the order of 100~Myr) could still be compatible with being remnants of past clustered star formation.

\subsection{Caveats and future outlook}
\label{caveat}
It is currently challenging to define whether an OC is “gravitationally bound.” In Section \ref{intro}, we mentioned that the criterion from \citetalias{HuntReffert2024} was based on their compactness, because contamination within measured velocity dispersions due to numerous factors prevented them from consistently applying the virial theorem for all their OCs (see their Section 2 for details). Further work is needed to better constrain the boundedness of OCs, such as through upcoming radial velocity surveys or methodological improvements to \citetalias{HuntReffert2024}'s Jacobi radius approach. Moreover, many OCs are still qualified as such in spite of being supervirial, even older ones. For instance, despite being $\sim$800~Myr old, and expected to dissolve within the next 30~Myrs, the Hyades is still referred as an OC \citep[e.g.,][]{OhEvans2020}.

On the one hand, residual gas expulsion occurs at ages where isochrone fitting lacks accuracy because young clusters are differentially reddened due to the effect of gas and dust, and therefore the change in boundedness for clusters younger than 10~Myr is hard to measure aside from in a few well-studied populations whose ages are precisely constrained \citep[e.g.,][]{CantatGaudinCasamiquela2024,RottensteinerMeingast2024}. On the other hand, there is a dearth of good theoretical models at the low-mass end of cluster formation that include feedback (e.g., simulations from \citealt{Ali2023} only include clusters with masses greater than 1000 M$_{\odot}$), although simulations from \citet{GonzalezSamaniego2020} show that feedback can cause a low-mass, self-gravitating OC to expand and disperse after $\sim$5--10 Myr.

Future research, such as extending the census of OB stars from \citetalias{Quintana2025} to greater distances (e.g., up to 2 kpc, Quintana et al. in prep.), will allow us to better estimate $\sum_{\rm SFR}$, and thus further constrain the nature of the environment in which stars form. Likewise, the next \textit{Gaia} data releases will be significant improved over \textit{Gaia} DR3 \citep{Brown2019}, enabling us to produce more accurate and complete catalogs of star clusters than ever before. Different regions of the Milky Way may have different cluster star formation fractions, and calculating this outside of the solar neighborhood will require us to derive a selection function for the OC census (\citealt{Hunt2025}, Hunt et al. in prep.) or for OB star censuses. A more precise $\sum_{\rm SFR}$ could tell us whether there remain statistically significant differences between the modes of star formation, accounting for the possibility that star formation could occur in both clustered and hierarchical environments. In any case, a comparison between the $\sum_{\rm SFR}$ derived from different types of objects is an effective way to constrain the fundamental nature of star formation that should be pursued further in the future.

\begin{acknowledgements}

We thank the anonymous referee for their insight that helped improving the manuscript. ALQ acknowledges a PSL fellowship granted by the Scientific Council from the Paris Observatory. The authors would like to thank Nick Wright for his thoughtful feedback, as well as Laia Casamiquela and Henny Lamers for insightful discussions. This paper makes uses of data processed by the Gaia Data Processing and Analysis Consortium (DPAC, https://www.cosmos.esa.int/web/gaia/dpac/consortium) and obtained by the Gaia mission from the European Space Agency (ESA) (https://www.cosmos.esa.int/gaia). This work also used \textit{TOPCAT} \citep{Topcat}, Astropy \citep{Astropy}, and the Vizier database (operated at CDS, Strasbourg, France). 
\end{acknowledgements}

\bibliographystyle{aa}
\bibliography{Bibliography} 

\end{document}